\documentclass[12pt]{article}
\usepackage{amssymb,amsmath,bm,epsfig}

\renewcommand{\thefootnote}{\fnsymbol{footnote}}
\newcommand{\etatwo}[2]{\sqrt2\left(\eta_{#1}-\eta_{#2}\right)}
\newcommand{\etafour}[4]{\eta_{#1}-\eta_{#2}-\eta_{#3}+\eta_{#4}}
\newcommand{\etafourp}[4]{\eta_{#1}+\eta_{#2}+\eta_{#3}+\eta_{#4}}
\begin{document}
\title{}

\title{
\begin{flushright}
\begin{minipage}{0.2\linewidth}
\normalsize
WU-HEP-14-06 \\
EPHOU-14-012 \\*[50pt]
\end{minipage}
\end{flushright}
{\Large \bf 
Gaussian Froggatt-Nielsen mechanism \\ on magnetized orbifolds \\*[20pt] } }

\author{
Hiroyuki~Abe$^{1,}$\footnote{
E-mail address: abe@waseda.jp}, \ 
Tatsuo~Kobayashi$^{2}$\footnote{
E-mail address: kobayashi@particle.sci.hokudai.ac.jp}, \
Keigo~Sumita$^{1,}$\footnote{
E-mail address: k.sumita@moegi.waseda.jp}
\\ and \ 
Yoshiyuki~Tatsuta$^{1,}$\footnote{
E-mail address: y\_tatsuta@akane.waseda.jp}\\*[20pt]
$^1${\it \normalsize 
Department of Physics, Waseda University, 
Tokyo 169-8555, Japan} \\
$^2${\it \normalsize 
Department of Physics, Hokkaido University, Sapporo 060-0810, Japan} \\*[50pt]}

\date{
\centerline{\small \bf Abstract}
\begin{minipage}{0.9\linewidth}
\medskip 
\medskip 
\small 
We study the structure of Yukawa matrices derived from the supersymmetric 
Yang-Mills theory on magnetized orbifolds, which can realize the 
observed quark and charged lepton mass ratios as well as 
the CKM mixing angles, even with a small number of tunable parameters 
and without any critical fine-tuning. 
As a reason behind this, we find that the obtained Yukawa matrices possess 
a Froggatt-Nielsen like structure with Gaussian hierarchies, which provides 
a suitable texture for them favored by the experimental data. 
%This model is realized by the non-SUSY background 
%and we also study the spectrum including heavy particles caused 
%by the magnetic fluxes. 
\end{minipage}} 

\begin{titlepage}
\maketitle
\thispagestyle{empty}
\clearpage
\tableofcontents
\thispagestyle{empty}
\end{titlepage}

\renewcommand{\thefootnote}{\arabic{footnote}}
\setcounter{footnote}{0}

\section{Introduction}
The standard model (SM) is a successful theory describing all the elementary 
particles discovered so far. 
However, it is still mysterious why it has such a 
complicated structure, e.g., the product gauge group, the three generations 
and the hierarchical Yukawa couplings required by observational results. 
Extra dimensional space would give 
an origin of the flavor structures, and that has been studied 
with a great variety. 
In particular, we focus on toroidal compactifications of higher-dimensional 
supersymmetric Yang-Mills (SYM) theories with magnetic fluxes on the torus, 
which induce the four-dimensional (4D) chiral spectra, like the SM 
\cite{Bachas:1995ik,Cremades:2004wa}. 

So far, some phenomenological studies of the higher-dimensional SYM theories 
with magnetized extra dimensions have been done. 
In the works, we constructed a phenomenological model \cite{Abe:2012fj} 
starting from the ten-dimensional (10D) SYM theory with a certain flux configuration, 
in which the magnetic fluxes on three factorizable two-dimensional (2D) 
tori give the semi-realistic flavor structures of the SM. 
Magnetic fluxes in extra dimensional space break the supersymmetry (SUSY) in general. 
The flux configuration studied in Ref. \cite{Abe:2012fj} preserves 
a $\mathcal N=1$ SUSY 
as a part of the $\mathcal N=4$ SUSY possessed by the 10D SYM theory 
with respect to the 4D supercharges, 
and we constructed the model within a framework of the low-scale SUSY breaking scenario. 
We have also searched the other flux configurations \cite{Abe:2013bba} 
and found that any other choice of factorizable magnetic fluxes 
\footnote{ We call fluxes factorizable if they do not cross over two tori. } 
would not induce plausible three generation structures 
even if we abandon preserving the $\mathcal N=1$ SUSY. 
Therefore the realistic three generation spectrum is uniquely determined to 
the one given in Ref. \cite{Abe:2012fj} from the 10D magnetized SYM 
on factorizable tori.

The situation changes if the chirality projections due to Yang-Mills fluxes 
are combined with those of the background geometry. 
The most simple but illustrating example to realize such a situation is 
magnetized orbifold models \cite{Abe:2008fi,Abe:2008sx,Abe:2009uz,Abe:2013bca}. 
The orbifold projections alter the relation between 
the magnitude of magnetic fluxes and the number of degenerate zero-modes 
identified with the number of generations. 
That is attractive because it provides a rich variety of 
phenomenological models obtained from the magnetized SYM theories. 
However, we found that any flux configurations on orbifolds yielding three 
generations of quarks and leptons do not preserve 
4D $\mathcal N=1$ SUSY \cite{Abe:2013bba}. 
%\footnote{We considered a orbifold projection 
%in the previous model \cite{Abe:2012fj}, but 
%that is only to eliminate extra massless modes and that has nothing to do with the three 
%generation structures.} 
That is, the magnetic fluxes on orbifold backgrounds break 
the SUSY corresponding to that of the MSSM at the compactification scale 
as far as those trigger the SM flavor structure. 
We think these facts are strong phenomenological predictions derived from 
the toroidal compactifications of 10D SYM theory. 
Therefore it would be important to study such a non-SUSY orbifold background, 
which can generate Yukawa structures different 
from those shown in Ref. \cite{Abe:2012fj}.

In this paper, we construct a magnetized orbifold model 
with a high-scale SUSY breaking caused by the magnetic fluxes on orbifolds. 
First, we focus on the structure of 2D orbifold 
because all the three generation structures of the SM should arise 
on a single 2D tori to get full rank Yukawa matrices \cite{Abe:2008sx} 
and that is adequate to discuss the Yukawa structures. 
We will identify a flux configuration whose predicted values 
of the quark and charged lepton masses 
and the CKM mixing angles \cite{Kobayashi:1973fv} are consistent 
with the experimental data. 
The resultant model is quite interesting and attractive because it has 
a smaller number of continuous free parameters 
(more strictly not parameters but vacuum expectation values(VEV) ) 
tunable to fit the experimental data, because the orbifold background 
does not allow nonvanishing continuous Wilson-lines.

Furthermore, we will find that the Yukawa matrices in this model 
have a reasonable structure to produce suitable hierarchies 
favored by the experimental data in wide parameter regions 
without fine-tuning. 
Finally, we give an example of embedding the six-dimensional (6D) model into 
the full 10D spacetime. 

\section{10D SYM with magnetized orbifolds}

\subsection{Zero-modes and Yukawa couplings}

We give a brief review of the 10D SYM theory with the magnetized and orbifold backgrounds. 
We start from the 
10D $U(N)$ SYM theory,
\begin{equation*}
S=\int d^{10}x \sqrt{-G}\left\{-\frac1{4g^2}{\rm tr}\left(F^{MN}F_{MN}\right)+
\frac i{2g^2}{\rm tr}\left(\bar \lambda \Gamma^M D_M\lambda\right)\right\}, 
\end{equation*} 
compactified on a product of 4D Minkowski space and three 
factorizable tori, $R^{3,1}\times (T^2)^3$. The covariant derivative $D_M$ and the field strength 
$F_{MN}$ contain the 10D vector field $A_M$ with $M,N =0,1,\cdots,9$ and $\lambda$ is the 10D 
Majorana-Weyl spinor field. The 10D metric $G_{MN}$ whose determinant is 
denoted by $G$ contains three torus metrics,
\begin{equation*}
g^{(i)}=\left(2\pi R^{(i)}\right)
\begin{pmatrix}
1& {\rm Re}\,\tau^{(i)} \\
{\rm Re}\,\tau^{(i)} & \left|\tau^{(i)}\right|^2 
\end{pmatrix},
\end{equation*} 
for $i=1,2,3$, 
where real parameters $R^{(i)}$ and complex parameters $\tau^{(i)}$ correspond to the sizes and 
shapes of the $i$-th torus $T^2$,  
respectively. 
The 10D fields are decomposed as 
\begin{eqnarray*}
\lambda\left(x^\mu,y^k\right)&=& \sum_{l,m,n} 
\chi_{lmn}\left( x^\mu\right)\otimes \psi^{(1)}_l\left(y^4,y^5\right)
\otimes \psi^{(2)}_m\left(y^6,y^7\right)\otimes \psi^{(3)}_n\left(y^8,y^9\right),\\
A_M\left(x^\mu,y^k\right)&=& \sum_{l,m,n} \varphi_{lmn,M}\left(x^\mu\right)
\otimes \phi^{(1)}_{l,M}\left(y^4,y^5\right)\otimes \phi^{(2)}_{m,M}\left(y^6,y^7\right)
\otimes \phi^{(3)}_{n,M}\left(y^8,y^9\right), 
\end{eqnarray*}
where $\psi^{(i)}_l\left(y^{2+2i},y^{3+2i}\right)$ is a 2D spinor 
corresponding to the  $l$-th mode on the $i$-th $T^2$. 
We concentrate on the zero modes $l=0$ in the following and denote them as 
\begin{equation*}
\psi^{(i)}_0=\begin{pmatrix}
\psi^{(i)}_+\\
\psi^{(i)}_-
\end{pmatrix}.
\end{equation*}

We introduce factorizable (Abelian) magnetic fluxes $F_{4,5}$, $F_{6,7}$ and $F_{8,9}$ on the tori, which have typically the following forms,
\begin{equation*}
F_{2+2i,3+2i}=\pi\begin{pmatrix}
M_1{\bm 1}_{N_1}&&\\
&\ddots&\\
&&M_n{\bm 1}_{N_n}
\end{pmatrix}, 
\end{equation*}  
where $M_k\in\mathbb{Z}$ because of the Dirac's quantization condition and 
${\bm 1}_{N_k}$ denotes the $(N_k \times N_k)$ unit matrix.
These fluxes break the $U(N)$ symmetry down to 
$U(N_1) \times \cdots \times U(N_n)$. 
Then, we denote matter fields with the bifundamental representation $(N_a, \bar N_b)$ under the unbroken gauge group 
$U(N_a)\times U(N_b)$ as ${\psi^{(i)}_\pm}^{ab}$. Using the gamma matrices
\begin{equation*}
\Gamma^{2+2i}=\begin{pmatrix}
0&1\\
1&0
\end{pmatrix},\hspace{3em}
\Gamma^{3+2i}=\begin{pmatrix}
0&-i\\
i&0
\end{pmatrix}, 
\end{equation*} 
the Dirac equations for ${\psi^{(i)}_\pm}^{ab}$ on the $i$-th $T^2$ 
are given by 
\begin{eqnarray} 
\left[\bar\partial_i+\frac{\pi M_{ab}^{(i)}}{2{\rm Im}\tau^{(i)}}z_i\right]
{\psi^{(i)}_+}^{ab}&=&0,\label{eq:posi}\\
\left[\partial_i-\frac{\pi M_{ba}^{(i)}}{2{\rm Im}\tau^{(i)}}\bar z_i\right]
{\psi^{(i)}_-}^{ab}&=&0,\label{eq:nega}
\end{eqnarray}
where $M_{ab}^{(i)}=M_a^{(i)}-M_b^{(i)}$, $z_i=y_{2+2i}+\tau^{(i)}y_{3+2i}$ 
and $\partial_i=\partial/\partial z_i$.

In Eq.~(\ref{eq:posi}), degenerate zero-modes can be obtained 
with the number of degeneracy 
$M_{ab}^{(i)}$ if $M_{ab}^{(i)}>0$ , and their wavefunctions are given by \cite{Cremades:2004wa}
\begin{eqnarray}
{\psi^{(i)}_+}^{ab, I} &=& \Theta^{I,M^{(i)}_{ab}}\left(z_i,\tau^{(i)}\right),\label{eq:posizero}
\end{eqnarray}
where 
\begin{eqnarray*}
\Theta^{I,M}\left(z,\tau\right) &=& \mathcal N_I \cdot e^{i \pi M z 
\mathrm{Im}\, z / \mathrm{Im}\, \tau} \cdot \vartheta \begin{bmatrix}
I/M\\[5pt] 0\end{bmatrix} (M z , M \tau ),\\
\vartheta \begin{bmatrix}a\\[5pt] b\end{bmatrix} (\nu, \tau) &=& \sum_{l \in \mathbb{Z}} 
e^{\pi i (a+l)^2 \tau} e^{2 \pi i (a+l) (\nu+b)},
\end{eqnarray*}
and normalization factors $\mathcal N_I$ are determined as 
\begin{equation}
\int_{T^2} d^2z~ \Theta^{I,M}\left(\Theta^{J,M}\right)^*=\delta_{IJ}.\label{eq:norma}
\end{equation}
The index $I$ in Eq.~(\ref{eq:posizero}) labels degenerate zero-modes, 
$I=0,~1,~\cdots ,M_{ab}^{(i)}-1$. 
In this case, the conjugate field, ${\psi_+^{(i)}}^{ba}$, has no zero-modes because of the negative fluxes, 
$M_{ba}^{(i)}<0$. 
As for the other equation (\ref{eq:nega}), ${\psi_-^{(i)}}^{ab}$ has $|M_{ab}^{(i)}|$ degenerate 
zero-modes and their wavefunctions are given by $\left(\Theta^{I,M}\right)^*$ if $M_{ab}^{(i)}<0$, 
and then ${\psi_-^{(i)}}^{ba}$ has no zero-modes. 
Note that, in the case with $M_{ab}^{(i)}=0$, each of ${\psi_+^{(i)}}^{ab}$, 
${\psi_+^{(i)}}^{ba}$, ${\psi_-^{(i)}}^{ab}$ and ${\psi_-^{(i)}}^{ba}$ 
has a single zero-mode and their wavefunctions are constant on the torus. 
The total degeneracy is given by 
\begin{equation*}
\prod_{k=1}^3\left|M_{ab}^{(k)}\right|,
\end{equation*}
except for the case with $M_{ab}^{(k)}=0$ and such degenerate zero-modes 
can be identified as the generations of the SM particles. 
The magnetic fluxes can give a three generation structure.

Furthermore, each wavefunction of degenerate zero-modes is localized 
at a different point from each other by magnetic fluxes and 
overlap integrals of them on the torus give 
their Yukawa coupling constants. 
The Yukawa matrices can be 
hierarchical depending on the flux configurations.

We show them up to complex phases and global factors including 
the 10D gauge coupling $g$ as 
\begin{eqnarray}
Y_{\mathcal {IJK}}&=&  \lambda^{(1)}_{I_1J_1K_1}\lambda^{(2)}_{I_2J_2K_2}\lambda^{(3)}_{I_3J_3K_3},
\label{eq:4dyukawa}\\
\lambda_{I_iJ_iK_i}^{(i)}&=& \int_{T^2} d^2z_i~~ \Theta^{I_i,M_{ab}^{(i)}}(z_i,\tau^{(i)})
\Theta^{J_i,M_{bc}^{(i)}}(z_i,\tau^{(i)})\left(\Theta^{K_i,M_{ac}^{(i)}}(z_i,\tau^{(i)})\right)^*, \label{eq:intzero}
\end{eqnarray}
where $\mathcal I=\left(I_1,I_2,I_3\right)$,  $\mathcal J=\left(J_1,J_2,J_3\right)$, 
and $\mathcal K=\left(K_1,K_2,K_3\right)$. 
Thus the Yukawa couplings are given by the overlaps of three Jacobi-theta 
functions $\vartheta$. 
Hereafter we often omit the index $i$ for the $i$-th torus, when it is clear 
that we concentrate on one $T^2$.

This overlap integral (\ref{eq:intzero}) can be performed analytically 
by using the decomposition, 
\begin{eqnarray}
& & \Theta^{I,M_{ab}}(z,\tau)\Theta^{J,M_{bc}}(z,\tau) \nonumber \\ 
& & ~~~~~=\frac{\mathcal N_I\mathcal N_J}{\mathcal N_K}
\sum_{m\in\mathbb{Z}_{M_{ac}}}\Theta^{I+J+M_{ab}m,~M_{ac}}(z,\tau)
\times\vartheta \begin{bmatrix}\frac{M_{bc}I-M_{ab}J+M_{ab}M_{bc}m}{M_{ab}M_{bc}M_{ac}}\\[5pt] 
0\end{bmatrix} (0 , \tau M_{ab}M_{bc}M_{ac}),  \nonumber
\end{eqnarray}
with the normalization (\ref{eq:norma}), and we obtain 
\begin{equation}
\lambda_{IJK}=\frac{\mathcal N_I\mathcal N_J}{\mathcal N_K}\sum_{m\in\mathbb{Z}_{M_{ac}}}
\vartheta \begin{bmatrix}\frac{M_{bc}I-M_{ab}J+M_{ab}M_{bc}m}{M_{ab}M_{bc}M_{ac}}\\[5pt] 
0\end{bmatrix} (0 , \tau M_{ab}M_{bc}M_{ac})\times \delta_{I+J+M_{ab}m, K}~~, 
\label{eq:formyukawa}
\end{equation} 
where $K$ is defined up to ${\rm mod} M_{ac}$.

Next, we consider $Z_2$ projections on the magnetized tori. 
For instance, we take the orbifold $T^2/Z_2$ with coordinates $(y_4, y_5)$ 
identified as $(y_4, y_5)\rightarrow(-y_4, -y_5)$ under 
the $Z_2$ transformation, 
and the fields contained in the higher-dimensional SYM theory 
are projected as 
\begin{eqnarray}
A_m(-y_4, -y_5)&=&-PA_m(y_4, y_5)P\hspace{10pt}{\rm for}\,\, m=4,5,\nonumber\\
A_m(-y_4, -y_5)&=&+PA_m(y_4, y_5)P\hspace{10pt}{\rm for}\,\, m\neq4,5,\nonumber\\
\psi_\pm(-y_4, -y_5)&=&\pm P\psi_\pm(y_4, y_5)P,\label{eq:orbifrans}
\end{eqnarray}
where $P$ is a projection operator ($P^2={\bm 1}$). 
On such an orbifold background, either 
even or odd zero-modes remain and the number of zero-modes generated by magnetic fluxes 
are reduced (or vanish). 
Even and odd wavefunctions on the magnetized background are found as follows  
\cite{Abe:2008fi}
\begin{eqnarray*}
\Theta_{\rm even}^{I,M}(z,\tau)&=&\frac 1{\sqrt2}\left(\Theta^{I,M}(z,\tau)+\Theta^{M-I,M}(z,\tau)\right),\\
\Theta_{\rm odd}^{I,M}(z,\tau)&=&\frac 1{\sqrt2}\left(\Theta^{I,M}(z,\tau)-\Theta^{M-I,M}(z,\tau)\right),
\end{eqnarray*}
where we use the relation, $\Theta^{I,M}(-z)=\Theta^{M-I,M}(z)$. 
The number of surviving zero-modes can be counted and that is shown in Table \ref{tb:numzero}.
\begin{table}[htb]
\begin{center}
\begin{tabular}{cccccccccccc}  \hline
$M$&0&1&2&3&4&5&6&7&8&9&10\\ \hline\hline
even&1&1&2&2&3&3&4&4&5&5&6\\
odd&0&0&0&1&1&2&2&3&3&4&4\\ \hline
\end{tabular}
\end{center}
\caption{The numbers of degenerate zero-modes for even and odd wavefunctions.}
\label{tb:numzero}
\end{table}
From the table, we find that several magnitudes of magnetic fluxes yield 
three degenerate zero-modes and then three generations, 
while only the number of flux $M=3$ could give three generations 
without orbifoldings.

\subsection{Supersymmetry} 
10D SYM theories have the $\mathcal N=4$ SUSY counted by 4D supercharges and that can be broken 
by magnetic fluxes. 
However, flux configurations satisfying a certain condition can preserve a part of that, 
the 4D $\mathcal N=1$ SUSY, and then we can derive the MSSM-like models from the 10D SYM theory. 
One of the relevant SUSY preserving conditions is satisfied in the 
following $D$-flat direction of the fluxed $U(1)$, 
\footnote{See Ref.~\cite{Abe:2012ya}, which gives the manifest description 
of the $\mathcal N=1$ SUSY in the 10D magnetized SYM theory. }

\begin{equation}
\frac1{\mathcal A^{(1)}}F_{45}+\frac1{\mathcal A^{(2)}}F_{67}+\frac1{\mathcal A^{(3)}}F_{89}=0,
\label{eq:susycond}
\end{equation}
where $\mathcal A^{(i)}$ is an area of the $i$-th $T^2$. 
This condition guarantees the $\mathcal N=1$ SUSY 
at the compactification scale (e.g. GUT scale), 
and at the same time restricts flux model buildings strongly. 
The $D$-term contributions to the soft SUSY breaking masses arise 
to all the scalars charged under the fluxed $U(1)$ 
if the magnetic fluxes do not 
satisfy the condition (\ref{eq:susycond}) 
\footnote{
Magnetic fluxes other than $F_{45}$, $F_{67}$ and $F_{89}$ 
are encoded in the $F$-term SUSY conditions, 
and those are assumed to be absent in this paper. }.

We have been trying to construct a model along the $D$-flat direction 
(\ref{eq:susycond}) on the toroidal background \cite{Abe:2013bba}. 
As a result, we find that the model proposed in Ref. \cite{Abe:2012fj} 
is the unique solution that has (semi-)realistic spectra 
compatible with a low scale SUSY breaking scenario. 
Note that, in the model, 
the three generation structure is obtained without orbifold projections 
and three-block magnetic fluxes break 
the $U(8)$ group down to $U(4)_C\times U(2)_L\times U(2)_R$, i.e., 
to the Pati-Salam gauge group, 
which is further broken by Wilson-lines. 
Therefore it has been confirmed that no more three-generation models are 
possible with factorizable fluxes satisfying Eq.~(\ref{eq:susycond}).

Here we remark that the continuous Wilson-lines are forbidden on orbifolds 
because the corresponding components of vector fields have $Z_2$ parity odd. 
%If the three generation structure is given as degenerate zero-modes reduced 
%by orbifold projections, Wilson-lines cannot 
%be introduced on the torus. 
Without Wilson-lines, the fluxes by themselves must further break 
$U(4)_C\times U(2)_L\times U(2)_R$ 
to make a difference between the quarks and the leptons. 
Then it requires more-than-three blocks structure of the flux matrices, 
with which the $D$-flat condition (\ref{eq:susycond}) is hardly satisfied 
with just three torus parameters $\mathcal A^{(i)}$ \cite{Abe:2013bba}. 
That is the reason why the magnetized orbifold models cannot be realized 
with the preserved $\mathcal N=1$ SUSY. 

On the other hand, if we abandon the low-scale SUSY in the model building 
and permit non-SUSY fluxes out of the $D$-flat direction (\ref{eq:susycond}), 
then various orbifold models are expected to arise at first glance. 
However, it is not easy to make such models phenomenologically viable, 
because Wilson-lines are not allowed on orbifolds as mentioned above, 
which played a role in Ref. [3] to fit the quark/lepton masses/mixings 
to their experimental values as well as 
to break the Pati-Salam gauge symmetry. 
Since Wilson-lines shift the localization points of 
the zero-mode wavefunctions, we can control the size of 
their overlaps and then order of the magnitude of their Yukawa couplings. 
That is, Wilson-lines are 
significant degrees of freedom to realize experimental data 
in the previous model without orbifoldings.

Based on the above discussions, in the following, we try to construct 
non-SUSY orbifold models with magnetic fluxes yielding realistic flavor 
structures out of the $D$-flat direction (\ref{eq:susycond}). 
We will also discuss the effects of $D$-term contributions to scalar masses 
of the order of the compactification scale, that leads to 
the high-scale SUSY breaking scenario.

\section{The model }\label{sec:mode}

Here we study model building and derive realistic quark and lepton mass matrices.

\subsection{Model construction}

In order to obtain full rank Yukawa matrices, three generations 
of both left- and right-handed matters must be originated 
from the same torus. 
%We are constructing a model, and then, it is significant  
%both of the three generation structures of the left-handed and 
%the right handed matters must be generated on the same torus 
%to get full rank Yukawa matrices at the tree-level. 
Configurations of magnetic fluxes on the other two tori have to 
be determined to leave the SM contents unchanged and 
to eliminate extra fields. 
As we declared in the previous section, they are no longer restricted 
by the SUSY condition (\ref{eq:susycond}), that is, 
we consider the high-scale SUSY breaking scenarios. 
%Then, the structure of the two tori has no effect 
%on the values of Yukawa couplings except for the 
%$\mathcal O(1)$ global constant factors. 
Since we require that the structure of the other two tori must not change 
or spoil the three-generation structure at all, 
$\lambda^{(2)}$ and $\lambda^{(3)}$ in Eq.~(\ref{eq:4dyukawa}) 
cannot carry the flavor index when we consider the case 
that the three generations are induced on the first $T^2$. 
Therefore 4D Yukawa couplings are represented by 
\begin{equation*}
Y_{IJK}=\lambda^{(1)}_{IJK}\lambda^{(2)}\lambda^{(3)},
\end{equation*} 
where $\lambda^{(2)}$ and  $\lambda^{(3)}$ are $\mathcal O (1)$ global factors. 

We restrict ourselves to the single torus for a while, 
where three degenerate zero-modes arise, 
and we discuss about the whole extra dimensional space later. 
As shown in Table \ref{tb:numzero}, three degenerate zero-modes are 
realized by four types of magnetic fluxes in orbifold models; 
$M=4,5$ for $Z_2$ even modes and $M=7,8$ for $Z_2$ odd modes. 
Then, magnetic fluxes of the Higgs sector are automatically determined 
as a consequence of the gauge symmetry and the other consistency 
if the fluxes of the left- and right-handed sectors are fixed. 
Twenty flux configurations, which induce the three-generation left- and 
right-handed matters and certain numbers of Higgs fields on a single torus, 
were listed in Ref.~\cite{Abe:2008sx}. 
%They also studied the quark masses and mixings numerically. 
%We consider both of the lepton and the quark sector 
%and embedding such models into 10D spacetime. 
In the following, we construct a model on the orbifold background 
that contains the three-generation quarks and leptons 
simultaneously realizing a (semi-) realistic patterns 
of the mass ratios and the CKM mixing angles \cite{Kobayashi:1973fv}, 
even without the Wilson-line parameters. 

We start from 10D U(8) SYM theory with the following flux configuration 
on the first 
$T^2$ ($y_4,y_5$-directions), 
\begin{equation}
F_{45}=2\pi
\begin{pmatrix}
0\times{\bm1}_{3}&&&\\
&1&&\\
&&5\times{\bm1}_2&\\
&&&-7\times{\bm1}_2
\end{pmatrix}. \label{eq:flux}
\end{equation}
We also have other plausible configurations of magnetic fluxes 
(summarized in Appendix \ref{sec:appa}). 
Among them, the configuration (\ref{eq:flux}) is simple and instructive 
to demonstrate a mechanism which is a main proposal of this paper. 
We concentrate on this one in this paper and the other models would 
be studied elsewhere.

The magnetic flux (\ref{eq:flux}) breaks the $U(8)$ group down to 
$SU(3)_C\times SU(2)_L\times SU(2)_R$ 
up to $U(1)$s and leads to the model shown in Table \ref{tb:model}. 
%In the way, $U(4)_C$ is broken down to $U(3)_C\times U(1)$ and 
%all the extra colored fields can be eliminated by the magnetic fluxes and orbifold projections, 
%we will show later. These two are assumed in the Ref.~\cite{Abe:2008sx}. 
%However, 
%There is the unbroken $SU(2)_R$ and it is difficult to break it  
%by the magnetic fluxes and/or orbifold projections in realistic models, 
%as shown later.
This table shows magnetic fluxes appearing in the quark and lepton sectors 
as well as the Higgs sector and also their $Z_2$ parities.
%To explain the situation, we show the magnetic fluxes felt by each of the sectors in Table \ref{tb:model}. 
\begin{table}[htb]
\begin{center}
\begin{tabular}{c|c|c|c} \hline
&Left-handed&Right-handed&Higgs\\ \hline
Quark &$-5$ (even)&$-7$ (odd)&$12$ (odd)\\ \hline
Lepton &$-4$ (even)&$-8$ (odd)&$12$ (odd)\\ \hline
\end{tabular}
\end{center}
\caption{Magnetic fluxes appearing in the quark and lepton sectors on the first $T^2$.}
\label{tb:model}
\end{table}
We know from Table \ref{tb:numzero} and \ref{tb:model} that 
the flux configuration gives the three generations of quarks and leptons 
and five generations of Higgs multiplets. 
The last column in Table \ref{tb:model} shows  
the magnetic fluxes and the $Z_2$ parity of the Higgs sector.

It is important that the data of Higgs sector is determined 
by the left and right sectors to obtain 
the nonvanishing Yukawa couplings. 
Such couplings should be $Z_2$-even quantities (otherwise vanish) 
and the sums of magnetic fluxes vanish  
because of the gauge invariance, $M_{ab}+M_{bc}+M_{ca}=0$. 
We require a difference between the flux configurations of 
quarks and leptons \cite{kunoh:waseda2012} 
because their flavor structures are somewhat different from each other. 
The two different configurations, however, have to lead to the same 
consistent Higgs sector obtaining the nonvanishing Yukawa couplings. 
%The configuration of Higgs sector has to be consistent with 
%those of the left and right sectors of both 
%the quarks and the leptons simultaneously, 
%while we require a difference between the configurations of quarks and leptons %because their flavor structures are somewhat different from each other. 

$SU(2)_R$ remains unbroken in this model 
and the breaking sector (or mechanism) other than the magnetic fluxes and 
orbifoldings (and Wilson-lines of course) is required. 
We can consider such an additional sector as extensions of our model. 
In this paper, we just assume that the $SU(2)_R$ is broken somehow and 
it gives the different VEVs to the up- and down-type Higgs fields, 
because we are focusing on the structures of Yukawa matrices here. 
Therefore, in this model, the four Yukawa matrices (for $U,D,N,E$) 
have different structures due to the configurations of 
the magnetic fluxes, orbifold projections and VEVs of Higgs fields.

We show all the wavefunctions given by this flux configuration 
in Table \ref{tb:qlwave}. 
\begin{table}[htb]
\renewcommand\arraystretch{1.25}
\begin{center}
\begin{tabular}{c|c|c|c|c|c} \hline
&Q&u,d&L&e,$\nu$&H\\ \hline
0 
&$\Theta^{0,5}$ 
&$\frac1{\sqrt2}\left(\Theta^{1,7}-\Theta^{6,7}\right)$
&$\Theta^{0,4}$ 
&$\frac1{\sqrt2}\left(\Theta^{1,8}-\Theta^{7,8}\right)$ 
&$\frac1{\sqrt2}\left(\Theta^{1,12}-\Theta^{11,12}\right)$ \\ \hline
1 
&$\frac1{\sqrt2}\left(\Theta^{1,5}+\Theta^{4,5}\right)$ 
&$\frac1{\sqrt2}\left(\Theta^{2,7}-\Theta^{5,7}\right)$ 
&$\frac1{\sqrt2}\left(\Theta^{1,4}+\Theta^{3,4}\right)$ 
&$\frac1{\sqrt2}\left(\Theta^{2,8}-\Theta^{6,8}\right)$ 
&$\frac1{\sqrt2}\left(\Theta^{2,12}-\Theta^{10,12}\right)$ \\ \hline
2 
&$\frac1{\sqrt2}\left(\Theta^{2,5}+\Theta^{3,5}\right)$ 
&$\frac1{\sqrt2}\left(\Theta^{3,7}-\Theta^{4,7}\right)$ 
&$\Theta^{2,4}$ 
&$\frac1{\sqrt2}\left(\Theta^{3,8}-\Theta^{5,8}\right)$
&$\frac1{\sqrt2}\left(\Theta^{3,12}-\Theta^{9,12}\right)$ \\ \hline
3&&&&&$\frac1{\sqrt2}\left(\Theta^{4,12}-\Theta^{8,12}\right)$ \\ \hline
4&&&&&$\frac1{\sqrt2}\left(\Theta^{5,12}-\Theta^{7,12}\right)$ \\ \hline
\end{tabular}

\end{center}
\caption{Wavefunctions of the quarks, leptons and Higgs fields.}
\label{tb:qlwave}
\end{table}
Based on this table, we can evaluate all the Yukawa couplings analytically. 
We define the following $\eta$-function for the simple description 
of $\vartheta$-function, which appears in Eq.~(\ref{eq:formyukawa}), 
\begin{equation}
\eta_N=\vartheta \begin{bmatrix}N/M\\[5pt]0\end{bmatrix} 
(0 , \tau M),\label{eq:eta}
\end{equation} 
where 
%we denote $\tau^{(1)}$ as $\tau$ simply and 
$M$ is a product of three fluxes, $5\times7\times12=420$ for quarks and 
$4\times8\times12=384$ for leptons and 
$\tau = \tau^{(1)}$ in this section. 

We have the five Higgs fields in up- and down-sectors respectively, 
and the Yukawa coupling terms are written by 
\begin{equation*}
Y_{IJK}H_KQ_{L_I}Q_{R_J}=\left(Y_{IJ0}H_0+Y_{IJ1}H_1
+Y_{IJ2}H_2+Y_{IJ3}H_3+Y_{IJ4}H_4\right)Q_{L_I}Q_{R_J}. 
\end{equation*} 
One linear combination of the five can be identified as the SM Higgs fields, 
and usual Yukawa matrices of the SM are given by 
a linear combination of the five Yukawa matrices. 
For the quark sector, the five Yukawa matrices are given with $\eta_N$ as follows \cite{Abe:2008sx} 
\begin{eqnarray*}
Y_0&=&\frac1{\sqrt2}
\begin{pmatrix}
\etatwo5{65}&\etatwo{185}{115}&\sqrt2(\eta_{55}+\eta_{125})\\
\etafour{173}{103}{187}{163}&\etafour{67}{137}{53}{17}&
\etafour{113}{43}{127}{197}\\
\etafour{79}{149}{19}{89}&\etafour{101}{31}{199}{151}&
\etafour{139}{209}{41}{29}\\
\end{pmatrix},\\
Y_1&=&\frac1{\sqrt2}
\begin{pmatrix}
\etatwo{170}{110}&\etatwo{10}{130}&\sqrt2(\eta_{50}+\eta_{190})\\
\etafour{2}{142}{58}{82}&\etafour{178}{38}{122}{158}&
\etafour{62}{202}{118}{22}\\
\etafour{166}{26}{194}{94}&\etafour{74}{206}{46}{94}&
\etafour{106}{34}{134}{146}\\
\end{pmatrix},\\
Y_2&=&\frac1{\sqrt2}
\begin{pmatrix}
\etatwo{75}{135}&\etatwo{165}{45}&\etatwo{15}{195}\\
\etafour{173}{33}{117}{93}&\etafour{3}{207}{123}{87}&
\etafour{183}{27}{57}{153}\\
\etafour{9}{201}{51}{81}&\etafour{171}{39}{129}{81}&
\etafour{69}{141}{111}{99}\\
\end{pmatrix},\\
Y_3&=&\frac1{\sqrt2}
\begin{pmatrix}
\etatwo{100}{140}&\etatwo{80}{200}&\etatwo{160}{20}\\
\etafour{68}{208}{128}{152}&\etafour{172}{32}{52}{88}&
\etafour{8}{148}{188}{92}\\
\etafour{184}{44}{124}{164}&\etafour{4}{136}{116}{164}&
\etafour{176}{104}{64}{76}\\
\end{pmatrix},\\
Y_4&=&\frac1{\sqrt2}
\begin{pmatrix}
\etatwo{145}{205}&\etatwo{95}{25}&\etatwo{85}{155}\\
\etafour{107}{37}{47}{23}&\etafour{73}{143}{193}{157}&
\etafour{167}{97}{13}{83}\\
\etafour{61}{131}{121}{11}&\etafour{179}{109}{59}{11}&
\etafour{1}{71}{181}{169}\\
\end{pmatrix}.
\end{eqnarray*}
Also, the Yukawa matrices of the charged leptons are given by \cite{Abe:2008sx}  
\begin{align*}
Y_0&=
\begin{pmatrix}
y_b&0&-y_l\\
0&\frac1{\sqrt2}(y_e-y_i)&0\\
-y_f&0&y_h
\end{pmatrix},&
Y_1&=
\begin{pmatrix}
0&y_c-y_k&0\\
\frac1{\sqrt2}(y_b-y_h)&0&\frac1{\sqrt2}(y_f-y_l)\\
0&0&0
\end{pmatrix},\\
Y_2&=
\begin{pmatrix}
-y_j&0&y_d\\
0&\frac1{\sqrt2}(y_a-y_m)&0\\
y_d&0&-y_j
\end{pmatrix},&
Y_3&=
\begin{pmatrix}
0&0&0\\
\frac1{\sqrt2}(y_f-y_l)&0&\frac1{\sqrt2}(y_b-y_h)\\
0&y_c-y_k&0
\end{pmatrix},\\
Y_4&=
\begin{pmatrix}
y_h&0&-y_f\\
0&\frac1{\sqrt2}(y_e-y_i)&0\\
-y_l&0&y_b
\end{pmatrix},
\end{align*}
where
\begin{align*}
y_a&=\etafourp{0}{96}{192}{96},&
y_b&=\etafourp{4}{100}{188}{92},\\
y_c&=\etafourp{8}{104}{184}{88},&
y_d&=\etafourp{12}{108}{180}{84},\\
y_e&=\etafourp{16}{112}{176}{80},&
y_f&=\etafourp{20}{116}{172}{76},\\
y_g&=\etafourp{24}{120}{168}{72},&
y_h&=\etafourp{28}{124}{164}{68},\\
y_i&=\etafourp{32}{128}{160}{64},&
y_j&=\etafourp{36}{132}{156}{60},\\
y_k&=\etafourp{40}{136}{152}{56},&
y_l&=\etafourp{44}{140}{148}{52},\\
y_m&=\etafourp{48}{144}{144}{48}.
\end{align*}

Note again that the different VEVs of up- and down-type 
Higgs yield the difference between up- and down-type Yukawa matrices. 
%Note that, a difference between the up/down sector is given by the 
%VEVs of the two-type Higgs fields, $H_u$ and $H_d$, again. 

\subsection{Gaussian Froggatt-Nielsen mechanism} 

The tunable continuous parameters relevant to the generation structure 
in the 6D model with the flux configuration (\ref{eq:flux}) 
are the complex structure of the torus $\tau^{(1)}$ and the VEVs 
$v_{ui} \equiv \langle H_{ui} \rangle$ and  
$v_{di} \equiv \langle H_{di} \rangle$ of the five pairs 
of up- and down-type Higgs fields $H_{ui}$ and $H_{di}$ ($i=0,1,\ldots,4$). 
In the following, these parameters are fixed 
\footnote{We do not treat complex phases of Yukawa couplings by setting 
${\rm Re}\,\tau^{(1)}=0$ and also the (generation-independent) global factors 
of them are assumed to be of ${\cal O}(1)$.} 
to fit the resultant quark and charged lepton mass ratios 
as well as the CKM mixing angles to the observed values. 
If we assume some nonperturbative and/or higher-order effects to generate 
Majorana mass terms for the right-handed neutrinos, the neutrino masses 
as well as the lepton mixing matrix can be also analyzed, 
which is beyond the scope of this paper.\footnote{See e.g. 
 \cite{Blumenhagen:2006xt}.}
We do not take renormalization group effects on Yukawa couplings 
into account, those are negligible and dependent to the details of 
the $SU(2)_R$ (as well as of the SUSY) breaking sector.

%We can estimate Yukawa matrices numerically depending on the values of 
%the Higgs VEVs and ${\rm Im}\,\tau^{(1)}$
%\footnote{A real part of $\tau^{(1)}$ gives the physical phases 
%and we do not study that in the paper. We ignore other some 
%parameters giving $\mathcal O(1)$ global factors. }. 
%We calculate the mass ratios of 
%the quarks and charged leptons and the CKM mixing angles at the 
%compactification (e.g.,GUT) scale. 
%If we also try to realize the tiny neutrino masses,  
%the heavy Majorana mass terms are required. 
%Although we can take account into the RGE flows, 
%that depends on models including the $SU(2)_R$ breaking sector 
%and RGE flows will not change the structures of the Yukawa matrices 
%so much, thus we study without them here. 
%For a while, we set ${\rm Im}\,\tau^{(1)}= 1.5$.
%There are five pairs of Higgs fields, $H_i$ with $i=0,\cdots,4$.
%We denote their VEVs by $v_{ui}$ ($v_{di}$) for the up-sector (down-sector) 
%with $i=0,\cdots,4$. 
We first study 
the structures of the Yukawa matrices analytically, 
and utilize $\eta_N$ with ${\rm Im}\,\tau^{(1)}= 1.5$, 
which is defined in Eq.~(\ref{eq:eta}), 
\begin{equation*}
\eta_N=\vartheta \begin{bmatrix}N/M\\[5pt]0\end{bmatrix} (0 , 1.5Mi)=
\sum_{l \in \mathbb{Z}} e^{-1.5\pi  (N^2/M+2Nl+M l^2)}, 
\end{equation*}
where $M=420$ for quarks and $M=384$ for leptons. 
For a large $M$ such as $M=\mathcal O(100)$, 
contributions from the higher modes $l\neq0$ are strongly suppressed 
and we can use the following approximation, 
\begin{equation} 
\eta_N\sim e^{-\frac{1.5\pi}M N^2}.\label{eq:etaap}
\end{equation}
It is obvious that $\eta_N$ is larger than $\eta_{N'}$ for $N<N'$, 
which allows us to rewrite the five Yukawa matrices for the quark sector as 
\begin{align}\label{eq:Y0-4}
Y_0&\sim
\begin{pmatrix}
\eta_{5}&-\eta_{115}&\eta_{55}\\
-\eta_{103}&\eta_{17}&-\eta_{43}\\
-\eta_{19}&-\eta_{31}&\eta_{29}
\end{pmatrix},&
Y_1&\sim
\begin{pmatrix}
-\eta_{110}&\eta_{10}&\eta_{50}\\
\eta_{2}&-\eta_{38}&\eta_{22}\\
-\eta_{26}&-\eta_{46}&-\eta_{34}
\end{pmatrix}, \nonumber \\
Y_2&\sim
\begin{pmatrix}
\eta_{75}&-\eta_{45}&\eta_{15}\\
-\eta_{33}&\eta_{3}&-\eta_{27}\\
\eta_{9}&-\eta_{39}&\eta_{69}
\end{pmatrix},&
Y_3&\sim
\begin{pmatrix}
\eta_{100}&\eta_{80}&-\eta_{20}\\
\eta_{68}&-\eta_{32}&\eta_{8}\\
-\eta_{44}&\eta_{4}&-\eta_{64}
\end{pmatrix}, \nonumber \\
Y_4&\sim
\begin{pmatrix}
\eta_{145}&-\eta_{25}&\eta_{85}\\
\eta_{23}&\eta_{73}&-\eta_{13}\\
\eta_{11}&\eta_{11}&\eta_{1}
\end{pmatrix},
\end{align}
up to $\mathcal O (1)$ factors.

Note that the $(Y_4)_{ij}$ matrix has the hierarchy 
\begin{equation}
(Y_4)_{ij} \ll (Y_4)_{k \ell},
\end{equation}
for $i \leq k$ and $j \leq \ell$, and the 
$(Y_3)_{ij}$ matrix has a similar hierarchy. 
(There are a few exceptional entries with the smaller values.) 
That is quite useful to realize both the hierarchical quark masses, 
$m_u \ll m_c \ll m_t$ and $m_d \ll m_s \ll m_b$, 
and small mixing angles observed by the experiments. 
On the other hand, the $(Y_0)_{ij}$ and $(Y_1)_{ij}$ matrices 
have the hierarchy opposite to the above.
In the $(Y_2)_{ij}$ matrix, the $(2,2)$ entry 
as well as the $(1,3)$ and $(3,1)$ entries are large compared with 
those in $(Y_3)_{ij}$ and $(Y_4)_{ij}$, 
and that would be useful to make the corresponding entries in the mass matrices 
larger.
Thus, the matrices, $(Y_3)_{ij}$ and $(Y_4)_{ij}$, as well as $(Y_2)_{ij}$ 
are interesting.
Indeed, we find that $Y_{3 + 4} \equiv Y_3 + Y_4$ is estimated as 
\begin{equation}
Y_{3 + 4} \sim 
\begin{pmatrix}
\eta_{100}&-\eta_{25}&-\eta_{20}\\
\eta_{23}&-\eta_{32}&\eta_{8}\\
\eta_{11}&\eta_{4}&\eta_{1}
\end{pmatrix}.
\label{eq:qyuka}
\end{equation}

Interestingly, this matrix can be approximated by 
\begin{equation*}
Y^{({\rm G})}_{ij} = e^{-c \left(a_i+b_j\right)^2}, 
\end{equation*}
with proper values of $a_i$ and $b_j$, where $a_i$ $(b_j)$ depends on only 
the left-handed (right-handed) flavors.
It looks like the Froggatt-Nielsen (FN) form \cite{Froggatt:1978nt} , i.e.,
\begin{eqnarray} 
Y^{({\rm FN})}_{ij} = e^{-c\left(a'_i+b'_j\right)},
\end{eqnarray} 
up to ${\cal O}(1)$ factors, but $Y^{({\rm G})}_{ij}$
is a Gaussian form for $a_i+b_j$.
In the FN form, $a'_i$ and $b'_j$ correspond to 
some kind of quantum numbers of left-handed and right-handed fermions, 
respectively.\footnote{
The FN form of Yukawa matrices can be derived, e.g. in 
flavor models with extra U(1) \cite{Ibanez:1994ig}, five-dimensional models with 
the warped background \cite{Gherghetta:2000qt}, 
and four-dimensional models with strong dynamics such 
as conformal dynamics \cite{Nelson:2000sn}.}
The FN form has been extensively 
studied, because  it is very useful to obtain 
realistic values of quark masses and mixing angles. 

For simplicity, let us concentrate on the $(2 \times 2)$ low right matrix,
\begin{eqnarray}
Y^{(\rm FM)}_{2-3}=\left(
\begin{array}{cc}
Y^{({\rm FN})}_{22}  & Y^{({\rm FN})}_{23} \\
Y^{({\rm FN})}_{32} & Y^{({\rm FN})}_{33}
\end{array}
\right),
\end{eqnarray}
where $Y^{({\rm FN})}_{22} \leq Y^{({\rm FN})}_{23, 32} \leq Y^{({\rm FN})}_{33}$.
In this case, the mass ratio $m_2/m_3$ is estimated by 
$m_2/m_3 \sim Y^{({\rm FN})}_{23}Y^{({\rm FN})}_{32}/(Y^{({\rm FN})}_{33})^2$, 
\footnote{Note that $Y^{({\rm FN})}_{22} Y^{({\rm FN})}_{33} \sim 
Y^{({\rm FN})}_{23}Y^{({\rm FN})}_{32}$.} 
and the (2,3) entry of the diagonalizing matrix $V_{ij}$ for the left-handed sector 
is estimated as $V_{23} \sim Y^{({\rm FN})}_{23}/ Y^{({\rm FN})}_{33}$.
Then, we can realize the quark mass ratios and mixing angles by 
choosing proper values of $a'_i$ and $b'_j$ for the up and down-sectors 
such that $Y^{({\rm FN})}_{23}/ Y^{({\rm FN})}_{33} \sim V_{cb} = 0.04$ for 
both the up and down-sectors and 
$Y^{({\rm FN})}_{23}Y^{({\rm FN})}_{32}/(Y^{({\rm FN})}_{33})^2 \sim V_{cb}Y^{({\rm FN})}_{32}/Y^{({\rm FN})}_{33}  \sim m_c/m_t \sim 0.007$ 
for the up-sector and 
$Y^{({\rm FN})}_{23}Y^{({\rm FN})}_{32}/(Y^{({\rm FN})}_{33})^2 \sim V_{cb}Y^{({\rm FN})}_{32}/Y^{({\rm FN})}_{33}  \sim m_s/m_b \sim 0.03$, 
that is $Y^{({\rm FN})}_{32}/Y^{({\rm FN})}_{33} \sim 0.2$ for the up-sector and 
$Y^{({\rm FN})}_{32}/Y^{({\rm FN})}_{33} \sim 1$ for the down-sector.
Indeed, the quark mass ratios and mixing angles are often parametrized by using the Cabibbo angle 
$\lambda =0.22$.
Then, the realistic values can be realized by the following FN form:
\begin{eqnarray}
Y^{({\rm FN})}_{2-3} \sim \left(
\begin{array}{cc}
\lambda^3 & \lambda^2 \\
\lambda & 1
\end{array}
\right),
\end{eqnarray}
for the up-sector and 
\begin{eqnarray}
Y^{({\rm FN})}_{2-3} \sim \lambda' \left(
\begin{array}{cc}
\lambda^2 & \lambda^2 \\
1 & 1
\end{array}
\right),
\end{eqnarray}
with a proper value of $\lambda'$  
for the down-sector, 
up to ${\cal O}(1)$ factors.
Similarly, we can obtain mass ratios and mixing angles including the first generations.

One of the differences  between $Y^{({\rm FN})}$ and $Y^{({\rm G})}$ is 
that 
\begin{eqnarray}
\frac{Y^{({\rm G})}_{ii} Y^{({\rm G})}_{33}}{Y^{({\rm G})}_{i3}Y^{({\rm G})}_{3i}} \ll 1,
\end{eqnarray}
for $i=1,2$, 
but this ratio is of ${\cal O}(1)$ in the FN form.
However, the (1,1) and (2,2) elements can be almost irrelevant 
to obtain the realistic values in the above explanation 
of the FN form.
Thus, the form $Y^{({\rm G})}$ could lead to a result similar to the FN form.

As mentioned before, our model has the $SU(2)_R$ gauge symmetry.
That is, the up-sector and down-sector have the same Yukawa couplings.
If the patterns of VEVs $\langle H_i \rangle$ for $i=0,\cdots,4$ are the same 
in the up-sector and down-sector, we could not derive nonvanishing mixing angles.
Thus, we are required to have VEV patterns different between the up-sector 
and down-sector to obtain nonvanishing mixing angles, 
and we assume that such $SU(2)_R$ breaking happens.
As mentioned above, the Yukawa couplings $Y_3$ and $Y_4$ would be useful to realize 
the realistic mass hierarchy, and also $Y_2$ would be useful to enhance the $(2,2)$ entry.
Hence, we assume that $H_3$ and $H_4$ in both the up and down-sectors develop their 
VEVs and also $H_2$ in only the down-sector develops its VEV.
We assume that $v_{u3} \sim v_{u4}$, $v_{d3} \sim v_{d4}$ and 
$v_{d2} \ll v_{d4}$.
Then, the up-sector quark mass matrix is written by 
\begin{eqnarray}
m^{(u)} \approx
\begin{pmatrix}
\eta_{100}v_{u3}&-\eta_{25}v_{u4}&-\eta_{20}v_{u3}\\
\eta_{23}v_{u4}&-\eta_{32}v_{u3}&\eta_{8}v_{u3}\\
\eta_{11}v_{u4}&\eta_{4}v_{u3}&\eta_{1} v_{u4}
\end{pmatrix}
= v_{u4}
\begin{pmatrix}
\eta_{100}\rho_{u} &-\eta_{25}&-\eta_{20}\rho_{u}\\
\eta_{23} &-\eta_{32}\rho_{u}&\eta_{8}\rho_{u}\\
\eta_{11} &\eta_{4}\rho_{u}&\eta_{1} 
\end{pmatrix}
,
\end{eqnarray}
where $\rho_{u} = v_{u3}/v_{u4}$.
Similarly, the down-sector quark mass matrix is written by 
\begin{eqnarray}
m^{(d)} & \approx &
\begin{pmatrix}
\eta_{100}v_{d3}&-\eta_{25}v_{d4}&\eta_{15}v_{d2} - \eta_{20}v_{d3} \\
\eta_{23}v_{d4}&\eta_3 v_{d2} -\eta_{32}v_{d3}&\eta_{8}v_{d3}\\
\eta_{9}v_{d2}+\eta_{11}v_{d4}&\eta_{4}v_{d3}&\eta_{1} v_{d4}
\end{pmatrix}\\
&=& v_{d4}
\begin{pmatrix}
\eta_{100}\rho_{d}&-\eta_{25}&\eta_{15}\rho'_{d} - \eta_{20}\rho_{d}\\
\eta_{23}&\eta_3 \rho'_d-\eta_{32}\rho_{d}&\eta_{8}\rho_{d}\\
\eta_{9}\rho'_d+\eta_{11}&\eta_{4}\rho_d&\eta_{1} 
\end{pmatrix},
\end{eqnarray}
where $\rho_d= v_{d3}/v_{d4}$ and $\rho'_{d} = v_{d2}/v_{d4}$.

First, let us concentrate on the $(2 \times 2)$ low right matrices, again.
When $\rho_{u} \sim \rho_d$, we would obtain the unrealistic relation, 
$m_c/m_t \sim m_s/m_b$ for a negligible value $\rho'_d$. 
That is, the term with  $\rho'_d$ must be dominant in the (2,2) entry 
of the down-sector.
Then, the mass ratios and mixing angle are written by 
\begin{eqnarray}
\frac{m_c}{m_t} \sim \eta_4 \eta_8 (\rho^{(u)})^2, \qquad \frac{m_s}{m_b} \sim \eta_3  \rho'_d, 
\qquad V_{cb} \sim \eta_8 (\rho_u - \rho_d).
\end{eqnarray}
Note that $\eta_1 \approx 1$ and $\eta_{32} = 1.0\times 10^{-5}$ for $\tau = 1.5i$.
In addition, we obtain $\eta_3 \sim 0.9$, $\eta_4 \sim 0.8$ and $\eta_8 \sim 0.5$.
Then, we can realize the realistic values of the mass ratios 
and mixing angles by $\rho'_d = {\cal O}(0.01) - {\cal O}(0.1)$, $\rho_u = {\cal O}(0.1)$ and 
$\rho_u - \rho_d = {\cal O}(0.01) - {\cal O}(0.1)$.

Similarly, we can discuss the other elements.
Suppose that $\eta_{15} \rho'_{d}$  is larger than $\eta_{20} \rho_d$.
Then, the mixing angle $V_{us}$ is estimated by 
\begin{eqnarray}
V_{us} \sim \frac{\eta_{15}}{\eta_{8}} \frac{\rho'_d}{\rho_d}.
\end{eqnarray}
Since $\eta_{8} \sim 0.5$ and $\eta_{15} \sim 0.08$, we can realize 
$V_{us} ={\cal O}(0.1)$ for $\rho'_d/\rho_d = {\cal O}(0.1) - {\cal O}(1)$.
Furthermore, because $\det (m^{(u)}) \sim (v_{u4})^3 \eta_{23} \eta_{25}$ 
and $\det (m^{(d)}) \sim (v_{d4})^3 
\eta_3 \eta_{11} \eta_{15} (\rho'_d)^2$, 
we find 
\begin{eqnarray}
\frac{m_u}{m_t} \sim \frac{\eta_{23} \eta_{25}}{\eta_{4} \eta_{8} (\rho_u)^2}, \qquad 
\frac{m_d}{m_b} \sim \eta_{11} \eta_{15} \rho'_d.
\end{eqnarray}
Since $\eta_{23} \sim 0.003$, $\eta_{25} \sim 0.0009$, $\eta_{11} \sim 0.3$, and $\eta_{15} \sim 0.08$,  
we can realize the realistic mass ratios by $(\rho_u)^2 ={\cal O}(0.1)$ 
and $\rho'_d = {\cal O}(0.1)$.
When we combine with the above estimation, 
there are good parameter regions for $\rho_u$ and $\rho'_d$, still.

We show an example.
We set 
\begin{eqnarray}
\rho_u = 0.29, \qquad \rho_d = 0.38, \qquad \rho'_d = 0.1.
\end{eqnarray}
Then, we obtain the mass ratios and mixing angles shown in Table \ref{tb:ckm}.
\begin{table}[t]
\begin{center}
\begin{tabular}{|c||c|c|} \hline
 & Sample values & Observed \\ \hline
$(m_u, m_c, m_t)/m_t$ & 
$(1.7 \times 10^{-5}, 5.7\times 10^{-3}, 1 )$ &
$(1.5 \times 10^{-5}, 7.5\times 10^{-3}, 1 )$  
 
\\ \hline
$(m_d, m_s, m_b)/m_b$ & 
$(2.0 \times 10^{-3}, 6.8 \times 10^{-2}, 1)$ & 
$(1.2 \times 10^{-3}, 2.3 \times 10^{-2}, 1)$ 
\\ \hline
$(m_e, m_\mu, m_\tau)/m_\tau$ & 
$(2.7 \times 10^{-4}, 5.9 \times 10^{-2}, 1)$ & 
$(2.9 \times 10^{-4}, 6.0 \times 10^{-2}, 1)$ 
\\ \hline \hline 
$|V_{\rm CKM}|$ & 
\begin{minipage}{0.3\linewidth}
\begin{eqnarray} 
\left( 
\begin{array}{ccc}
0.96 & 0.29 & 0.01 \\
0.29 & 0.96 & 0.07 \\
0.01 & 0.07 & 1.0 
\end{array}
\right) 
\nonumber
\end{eqnarray} \\*[-20pt]
\end{minipage}
& 
\begin{minipage}{0.3\linewidth}
\begin{eqnarray} 
\left( 
\begin{array}{ccc}
0.97 & 0.23 & 0.0035 \\
0.23 & 0.97 & 0.041 \\
0.0087 & 0.040 & 1.0 
\end{array}
\right) 
\nonumber
\end{eqnarray} \\*[-20pt]
\end{minipage} \\ \hline
\end{tabular}
\end{center}
\caption{
The mass ratios of the quarks and the charged leptons, 
$(m_u,\, m_c,\, m_t)/m_t$, $(m_d,\, m_s,\, m_b)/m_b$ and 
$(m_e,\, m_\mu,\, m_\tau)/m_\tau$, 
and the absolute values of the CKM matrix $V_{\rm CKM}$ elements. 
The experimental data 
are quoted from Ref.~\cite{Beringer:1900zz}.} 
\label{tb:ckm}
\end{table}
To obtain these results, we have used the full formula without 
the approximation (\ref{eq:etaap}) and (\ref{eq:Y0-4}).
Thus our model can lead to a (semi-)realistic pattern.
The specific values of input parameters are not so important 
and such (semi-)realistic patterns can be realized 
in wide regions of the parameter space as far as 
the VEVs are similar values. 
That means the experimentally observed hierarchies 
in the quark sector are realized without any critical fine-tuning.

%%%%%%%%%%%%%%%%%%%%%%%%%%%%%%%%%%%%%%%%%%%%%%%%%%%%%%%%%%%

After the relevant parameters (VEVs) are fixed to fit 
the quark sector so far, we can evaluate the Yukawa matrices 
in the charged lepton sector as 
\begin{align*}
Y_0&\sim
\begin{pmatrix}
\eta_{4}&0&-\eta_{44}\\
0&\eta_{16}&0\\
-\eta_{20}&0&\eta_{28}
\end{pmatrix},&
Y_1&\sim
\begin{pmatrix}
0&\eta_{8}&0\\
\eta_{4}&0&\eta_{20}\\
0&0&0
\end{pmatrix},\\
Y_2&\sim
\begin{pmatrix}
-\eta_{36}&0&\eta_{12}\\
0&\eta_{0}&0\\
\eta_{12}&0&-\eta_{36}
\end{pmatrix},&
Y_3&\sim
\begin{pmatrix}
0&0&0\\
\eta_{20}&0&\eta_{4}\\
0&\eta_{8}&0
\end{pmatrix},\\
Y_4&\sim
\begin{pmatrix}
\eta_{28}&0&-\eta_{20}\\
0&\eta_{16}&0\\
-\eta_{44}&0&\eta_{4}
\end{pmatrix}.
\end{align*}
Then, we find the following charged lepton mass matrix, 
\begin{align*}
%y_{\rm lepton}&\sim
 m^{(\ell)}\sim  v_{d4}
\begin{pmatrix}
\eta_{28}&0&\eta_{12}\rho'_d - \eta_{20}\\
\eta_{20} \rho_d& \rho'_d+\eta_{16}&\eta_{4}\rho_d\\
\eta_{12} \rho'_d -\eta_{44}&\eta_{8}\rho_d&\eta_{4}
\end{pmatrix}. 
\end{align*} 
The numerical results of the charged lepton mass ratios are also shown 
in Table \ref{tb:ckm}.

We have shown analytically that the precise values of the complex structure 
and the ratios of Higgs VEVs specified above 
are not so critical to produce the observed hierarchies thanks to 
the nice texture of the Yukawa matrices obtained in the magnetized orbifold. 
Next, we numerically confirm that 
the suitable spectra can be given 
in wide parameter regions in terms of the Higgs VEVs and 
the complex structure ${\rm Im}\,\tau^{(1)}$.

\begin{figure}[htbp]
\begin{center}
\includegraphics[width=0.47\linewidth]{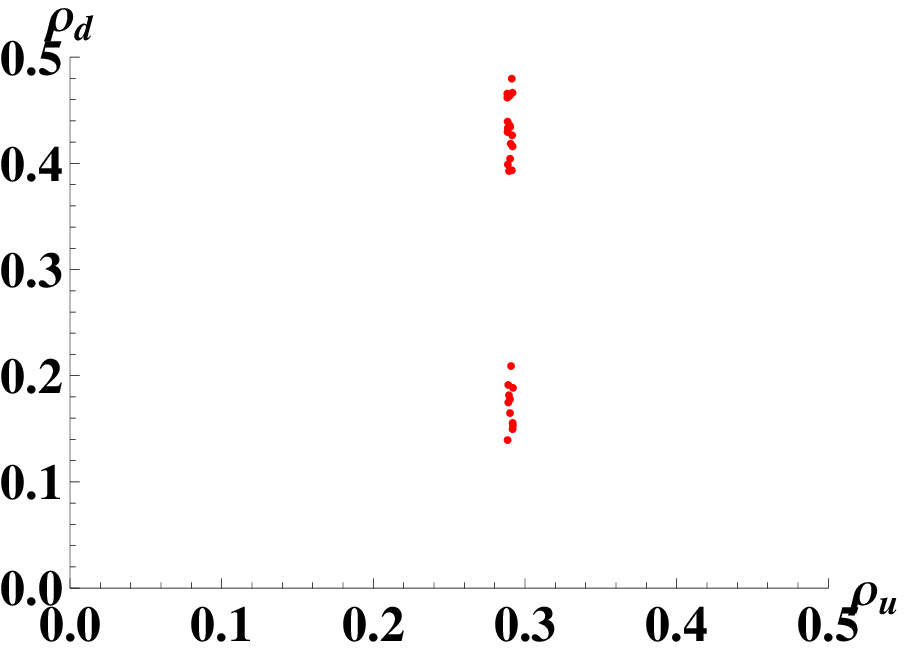}\\ \ \\
\includegraphics[width=0.47\linewidth]{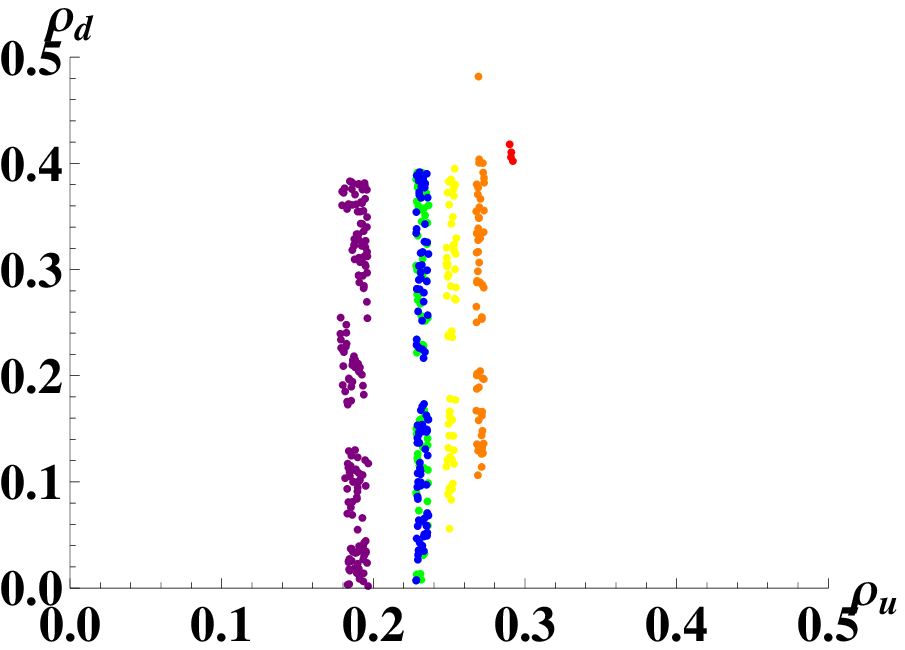}\\ \ \\
\includegraphics[width=0.47\linewidth]{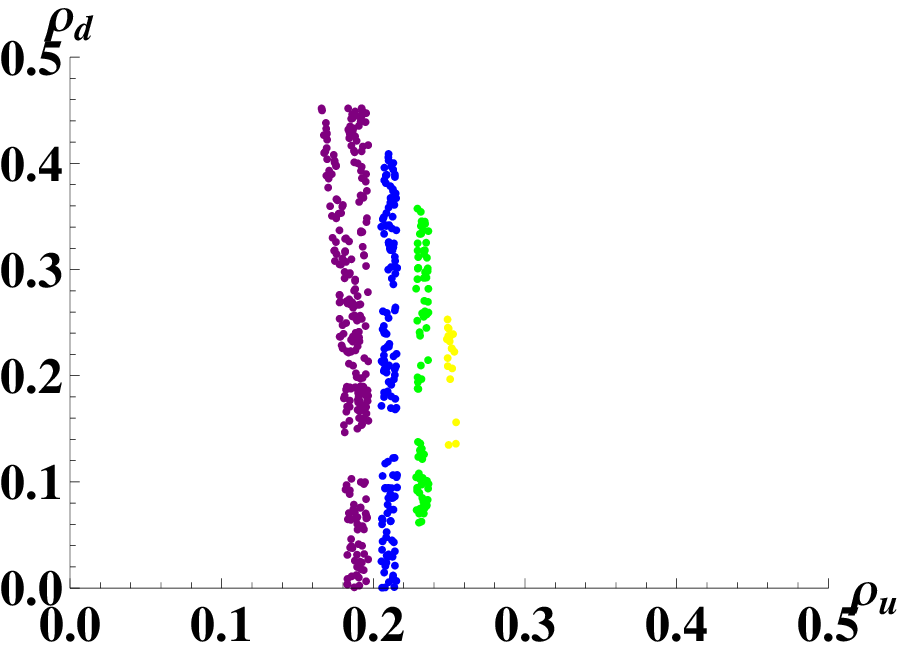}
\end{center}
\caption{
The dots represent the input values of ($\rho_{u},~\rho_{d}$) which yield 
the theoretical values of the mass ratios and the mixing angles 
within the range $0.2 \leq {\rm th/ex}\leq 5.0$. 
We choose values of $\rho_{u}$ and $\rho_{d}$ from 0 to 0.5 randomly 
with $N=10^4$ trials for each combination 
$(\rho'_{d} , {\rm Im}\, \tau^{(1)})$ with 
$\rho'_{d}=0.01~,0.05,~0.1$ respectively from the top panel 
and ${\rm Im} \,\tau^{(1)} =1.5,~1.6,~1.7,~1.8,~1.9,~2.0$ 
distinguished by six colors, red, orange, yellow, green, blue and violet 
respectively. The suitable hierarchies are realized in wide regions. 
 }
\label{fig:22}
\end{figure}

\begin{figure}[htbp]
\begin{center}
\includegraphics[width=0.47\linewidth]{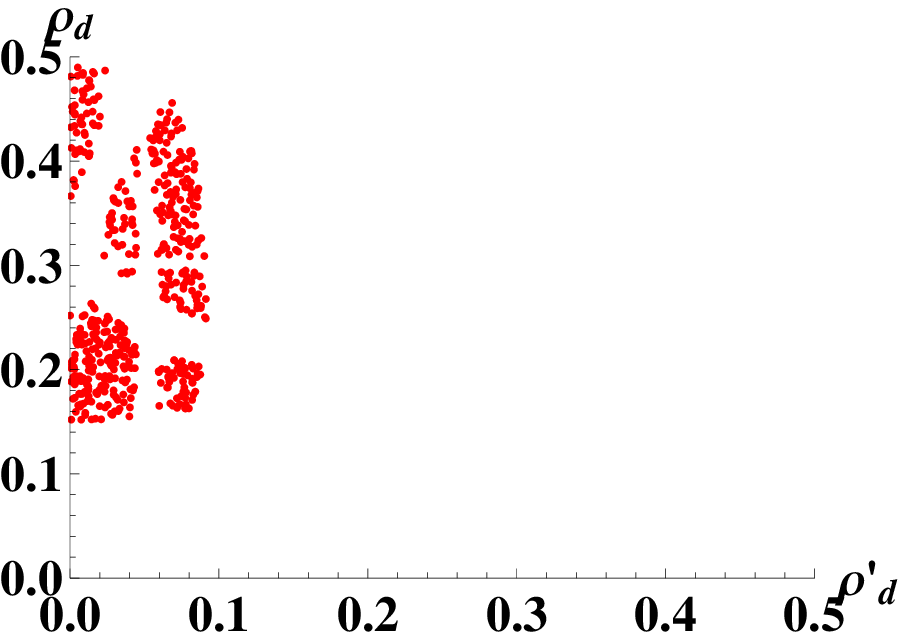}\\ \ \\
\includegraphics[width=0.47\linewidth]{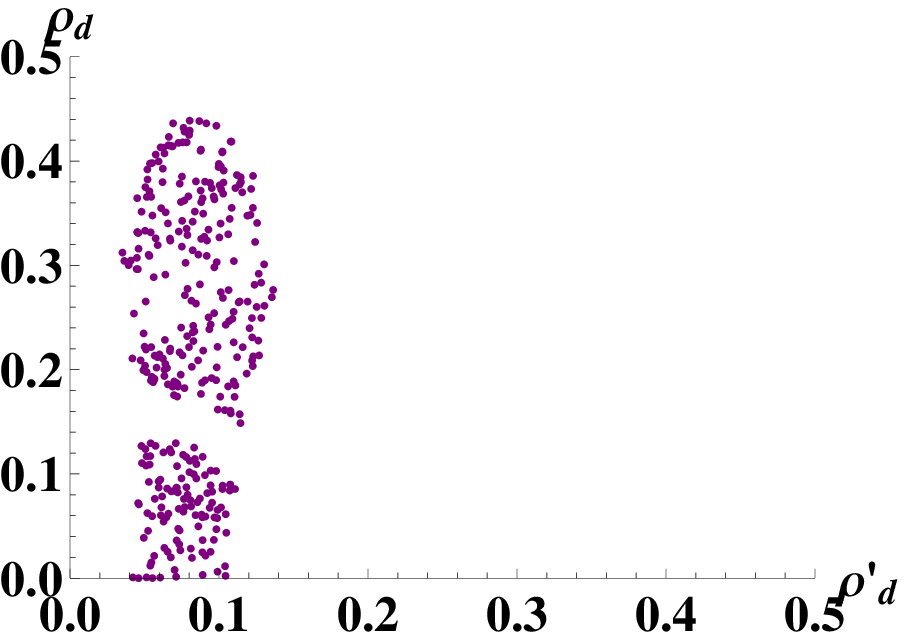}
\end{center}
\caption{
The dots represent the input values of ($\rho'_{d},~\rho_{d}$) which yield 
the theoretical values of the mass ratios and the mixing angles 
within the range $0.2 \leq {\rm th/ex}\leq 5.0$. 
We choose values of $\rho_{d}$ and $\rho'_{d}$ from 0 to 0.5 randomly 
with $N=10^4$ trials for each combination of 
$(\rho_u, {\rm Im}\,\tau^{(1)})=(0.3, 1.45)$(the upper panel) 
and $(\rho_u, {\rm Im}\,\tau^{(1)})=(0.2, 1.95)$(the lower panel). }
\label{fig:33}
\end{figure}

We show the stability of the input parameters numerically 
in Figs.~\ref{fig:22} and \ref{fig:33}. 
The two axes of each panel in Fig.~\ref{fig:22} represent 
$\rho_{u}$ and $\rho_{d}$, 
which are chosen randomly from 0 to 0.5. 
The input values indicated by each (colored) dot in the parameter space 
($\rho_u,~\rho_d$) yield 
the ratios of the theoretical values to the experimental center values 
(th/ex ratio) within a range 0.2 to 5.0 inclusively 
, with respect to the six mass ratios and the nine 
elements of the CKM matrix simultaneously. 
Three panels correspond to the three cases with 
$\rho'_{d} = 0.01,~0.05,~0.1$ respectively from the top panel, 
and the other Higgs VEVs are vanishing. 
We use the six colors, red, orange, yellow, green, blue and violet, of dots to distinguish 
${\rm Im}\,\tau^{(1)}=1.5,~1.6,~1.7,~1.8,~1.9,~2.0$ respectively. 
For each combination ($\rho'_d,~{\rm Im}\,\tau^{(1)}$), we try to dot 
$10^4$ times.
When ${\rm Im}\tau^{(1)}=1.4,~2.1$, 
the th/ex ratios are out of the range and no dot appears. 
We can see from this figure that the allowed values of each parameter 
are distributed over $\mathcal O(0.1)$ range of widths. In the parameter 
space with respect to VEVs (GeV) of the five Higgs fields, the allowed regions 
exist over $\mathcal O (1)$ or wider range of widths with 
$\tan\beta= v_u^{MSSM}/v_d^{MSSM}=\sqrt{(v_{u3}^2+v_{u4}^2)/(v_{d2}^2+v_{d3}^2+v_{d4}^2)} =1$, 
and even in the case with $\tan\beta =50$, the values of the most restricted VEVs can 
vary in  $\mathcal O (0.1)$ range. 
That is, the suitable hierarchies, 
like the one shown in Table \ref{tb:ckm}, are realized in wide 
regions of the parameter space. 
We also show explicitly that the allowed values of $\rho'_{d}$ exist 
over a wide range. In Fig.~\ref{fig:33}, 
the two axes of each panel represent $\rho'_{d}$ and $\rho_{d}$, 
which are chosen randomly from 0 to 0.5 with $10^4$ trials, 
and we draw a dot when the values of th/ex ratios are in the range 
$0.2\leq{\rm th/ex}\leq5.0$ as well as in Fig.~\ref{fig:22}. 
The other parameters are fixed as $(\rho_u, {\rm Im}\,\tau^{(1)})=(0.3, 1.45)$ 
in the upper panel and $(\rho_u, {\rm Im}\,\tau^{(1)})=(0.2, 1.95)$ 
in the lower panel, and the other Higgs VEVs are vanishing. 
From these figures, 
the obtained nice texture of the Yukawa matrices give 
the suitable hierarchies in wide regions of 
the parameter space without a critical fine-tuning.

\subsection{The 10D embedding and D-term contributions} 

We show an example of embedding our model into the 10D SYM 
theory with two independent $Z_2$ twists. 
We consider the magnetic fluxes on the second and the third tori as 
\begin{eqnarray}
F_{67}=2\pi\begin{pmatrix}
0\times{\bm 1}_4&0\\
0&-{\bm 1}_4
\end{pmatrix},\hspace{20pt}
F_{89}=2\pi\begin{pmatrix}
0\times{\bm 1}_4&0\\
0&{\bm 1}_4
\end{pmatrix}\label{eq:flux23}.
\end{eqnarray}
The SUSY condition (\ref{eq:susycond}) cannot be satisfied 
with the fluxes (\ref{eq:flux}) and (\ref{eq:flux23}). We also consider 
the 10D orbifold $R^{3,1}\times T^6/(Z_2\times Z_2')$. 
Under the $Z_2$ twist, the six toroidal coordinates transform as follows,
\begin{equation*}
\left(y_{4},y_{5},y_{6},y_{7},y_{8},y_{9}\right)\rightarrow
\left(-y_{4},-y_{5},-y_{6},-y_{7},y_{8},y_{9}\right),
\end{equation*}
and then, the transformations of the fields are already mentioned in Eq.~(\ref{eq:orbifrans}) 
with the projection operator 
\begin{equation}
P=\begin{pmatrix}
-{\bm 1}_4&0\\
0&{\bm 1}_4
\end{pmatrix}\label{eq:proj}. 
\end{equation}
The other, $Z_2'$ is 
\begin{equation} 
\left(y_{4},y_{5},y_{6},y_{7},y_{8},y_{9}\right)\rightarrow
\left(y_{4},y_{5},-y_{6},-y_{7},-y_{8},-y_{9}\right)\label{eq:secondz}
\end{equation}
with the same projection operator as that of $Z_2$.

This background is consistent with the parity assignment on the first torus 
shown in Table~\ref{tb:model} and 
does not change the flavor structure caused by the magnetic fluxes (\ref{eq:flux}). 
We summarize all the contents induced by this background in Table~\ref{tb:spectr}. 

\begin{table}[htb]
\begin{center}
\begin{tabular}{c|c|c|c}  \hline
& Fermions & Bosons&assignment\\\hline
SM gauge fields& massless&massless&$A_\mu,\lambda_{+++}$\\\hline
$Q$&3 generations&heavy&$A_6,A_7,\lambda_{-+-}$\\ 
$u,d$&3 generations&heavy&$A_8,A_9,\lambda_{--+}$\\ 
$L$&3 generations&heavy&$A_6,A_7,\lambda_{-+-}$\\ 
$\nu,e$&3 generations&heavy&$A_8,A_9,\lambda_{--+}$\\ \hline
$H_u,H_d$& massless&5 generations & $A_4,A_5,\lambda_{+--}$\\ \hline
Others & none& none&-\\ \hline
\end{tabular}
\end{center}
\caption{All the contents of the model. The four types of $\lambda_{\pm\pm\pm}$ 
are the 4D Weyl spinors contained in the 10D Majorana-Weyl spinor, and each sign represents 
chiralities on the three tori. Others(i.e, $\lambda_{---}$) would never appear 
because of the 10D Majorana condition. }
\label{tb:spectr}
\end{table}
We obtain all the SM contents. 
Amazingly, there are no extra fields at all,  
such as exotic modes, vector-like matters and diagonal adjoint fields, so-called ``open string moduli". 
The existence of such extra massless fields is known as a notorious open problem 
in string phenomenology. 
Indeed, there necessarily exist some extra massless fields in our past works. 
In this model, the two orbifold twists and the magnetic fluxes free from the SUSY conditions 
eliminate them completely.

Finally, we estimate the spectrum of scalar particles caused by the non-SUSY background~\footnote{As for the massless gauginos and higgsinos, 
we expect other contributions from moduli, anomaly and gauge mediated 
SUSY breaking as well as some SUSY mass terms. }. 
Masses of scalar fields on a magnetized torus are given by \cite{Cremades:2004wa}
\begin{equation*}
M^2_{\rm scalar}=2\pi\frac{|M_{ab}|}{\mathcal A},
\end{equation*}
and also, those of vector fields are 
\begin{equation*}
M^2_{\pm}=2\pi\frac{|M_{ab}|}{\mathcal A}\pm4\pi\frac{M_{ab}}{\mathcal A},
\end{equation*}
where $\pm$ represent mass eigenstates instead of the original 
2D real vector (e.g.,  $(A_4, A_5)$). 
Each of $A_m$ behaves as a vector field on the corresponding torus 
as well as a scalar field on the other two tori. 
The masses in the 4D effective field theory are given by 
the sums of their eigenvalues on the three tori. 
Accordingly, we can estimate the heavy scalar spectrum, i.e. squarks and sleptons,  as follows, 
\begin{eqnarray*}
m^2_{\tilde Q}&=&2\pi\left(\frac5{\mathcal A^{(1)}}+
\frac1{\mathcal A^{(2)}}\pm2\frac1{\mathcal A^{(2)}}+\frac1{\mathcal A^{(3)}}\right),\\
m^2_{\tilde L}&=&2\pi\left(\frac4{\mathcal A^{(1)}}+
\frac1{\mathcal A^{(2)}}\pm2\frac1{\mathcal A^{(2)}}+\frac1{\mathcal A^{(3)}}\right),\\
m^2_{\tilde u,\tilde d}&=&2\pi\left(\frac7{\mathcal A^{(1)}}+
\frac1{\mathcal A^{(2)}}+\frac1{\mathcal A^{(3)}}\pm2\frac1{\mathcal A^{(3)}}\right),\\
m^2_{\tilde e,\tilde \nu}&=&2\pi\left(\frac8{\mathcal A^{(1)}}+
\frac1{\mathcal A^{(2)}}+\frac1{\mathcal A^{(3)}}\pm2\frac1{\mathcal A^{(3)}}\right),
\end{eqnarray*} 
and the contribution to the SM Higgs fields is 
\begin{equation*}
m^2_H=2\pi\left(\frac{12}{\mathcal A^{(1)}}\pm2\frac{12}{\mathcal A^{(1)}}\right).
\end{equation*} 

Therefore, the lightest scalar mode in the Higgs sector 
has a tachyonic mass of the order of the compactification scale. 
We find that, in any models with various 10D embeddings eliminating all the extra massless fields, 
tachyonic modes arise inevitably. 
We have to assume additional perturbative contributions to the soft scalar 
masses, and even some nonperturbative effects or higher 
dimensional operators to give them enough masses to cancel these tachyonic 
contributions. 
For example, other SUSY breaking contributions mediated by 
moduli, anomaly and gauge interactions are able to cancel 
the tachyonic $D$-term contributions. 
Another example is Wilson lines. 
It is known that Wilson lines without magnetic fluxes arise 
massive modes \cite{Hamada:2012wj}. 
In the above example of 10D embedding, 
tachyonic Higgs fields have no fluxes on the two tori and 
a certain mass can be given by the Wilson lines. 
However, we have to remove the $Z_2'$ projection (\ref{eq:secondz}) 
to introduce the Wilson lines. 
Even in such a case, the (MS)SM contents shown in Table \ref{tb:spectr} 
are unchanged, but there also remain some massless Wilson line moduli. 
We would study more about the stability of our 10D embeddings elsewhere.

\section{Conclusions} 
We have realized a new-type of Yukawa structures in magnetized models 
with $Z_2$ orbifold twists. 
It is known that magnetic fluxes on a torus induce 
the chiral spectra and give them the degenerate (generation) structures. 
The magnitude of magnetic fluxes determine the number of 
generations, then, the orbifold projections give a variety of the 
model building with the three generations. 
We found such magnetized orbifold 
backgrounds do not preserve the 4D $\mathcal N=1$ SUSY 
\footnote{It is as far as we try to realize differences of Yukawa matrices between 
the quark and lepton sector at the tree-level.}, 
so in this paper, we have constructed a magnetized orbifold model 
with a non-SUSY background.

In Sec.~\ref{sec:mode}, we have constructed the model, which 
can realize the experimental data of the quarks and the charged leptons simultaneously. 
For the purpose, we have chosen the different magnetic fluxes between
the quark sector and the lepton sector under the condition that
the down-sector quarks and the charged leptons couple with
the same Higgs sector. 
Although the number of tunable parameters (VEVs) is extremely small in this model, we have been able to obtain the theoretical values of the masses and mixing angles, 
which is consistent with the experimental data, without fine-tunings. 
There is a reliable reason behind that, which has been revealed by 
our analysis of the Yukawa matrices.

The Yukawa matrix has the FN-like form with Gaussian hierarchies. 
The magnetic fluxes on the orbifold determine the 
``flavor quantum numbers'' and 
give the new type of texture to the Yukawa matrices. 
We can get the theoretical values favored by the experiments 
in the wide parameter regions thanks to this new texture 
obtained from the wavefunction localization on the magnetized orbifold. 
We have also confirmed these analytic results numerically in the figures. 
We would expect such a FN-like Yukawa structure with Gaussian hierarchies 
inspires various phenomenological studies.

Finally, we have given a sample of embedding the above generation structure 
into the product of three tori in 10D spacetime, 
and studied the spectrum caused by the SUSY breaking magnetic fluxes. 
The magnetic fluxes and the orbifold twists can eliminate 
all the extra massless fields in this model, 
although some scalar fields will become tachyonic. 
Such a model free from extra massless fields has never realized on SUSY magnetized backgrounds.

\subsection*{Acknowledgement}
H.A. and K.S. thank H. Kunoh for stimulating discussions 
in the early stage of this work. 
Y.T. would like to thank H. Ohki and Y. Yamada for useful discussions. 
H.A. was supported in part by the Grant-in-Aid for Scientific Research 
No.~25800158 from the Ministry of Education, Culture, Sports, Science 
and Technology (MEXT) in Japan. 
T.K. was supported in part by the Grant-in-Aid for Scientific Research 
No.~25400252 from the MEXT in Japan. 
K.S. was supported in part by a Grant-in-Aid for JSPS Fellows 
No.~25$\cdot$4968 and a Grant for Excellent Graduate
Schools from the MEXT in Japan. 
Y.T. was supported in part by a Grant for Excellent Graduate
Schools from the MEXT in Japan. 

\appendix

\section{Other flux configurations}
\label{sec:appa}
We require that the magnetic fluxes in the 
quark sector are different from ones in the lepton sector, 
but the quarks and leptons couple with the same Higgs sector. 
The configuration in Table \ref{tb:model} satisfies this condition. 
Obviously, when we exchange the magnetic fluxes 
between the quark and lepton sectors (the left-handed and right-handed 
sectors), that also satisfies the condition. 
In addition, we have found the other configurations 
satisfying the condition. 

The first one is that 
$M=-5$, $8$, and $-3$ for the left-handed quark, right-handed quark and 
Higgs sectors, and 
$M=-4$, $7$, and $-3$ for the left-handed lepton, right-handed lepton 
and Higgs sectors including the exchange of magnetic fluxes 
between the quark and lepton sectors (the left-handed and right-handed 
sectors). 
These lead to one-pair Higgs doublets just like the MSSM, but 
the Yukawa matrices do not have the Gaussian Froggatt-Nielsen 
texture. (We can easily see that they are phenomenologically 
disfavored.)

The second has a five-block structure breaking the $SU(2)_R$ symmetry. 
That is shown in Table \ref{tb:five} or with the quark/lepton exchange. 
\begin{table}[htb]
\begin{center}
\begin{tabular}{|c|c|c|c|c|c|c|c|} \hline
$Q$&$L$&$U$&$D$&$N$&$E$&$H_u$&$H_d$\\ \hline
 $-4$ (even) & $-5$ (even) & $-5$ (even) 
& $-8$ (odd) & $-4$ (even) & $-7$ (odd) &$9$ (even)&$12$ (odd)\\ \hline
\end{tabular}
\end{center}
\caption{Magnetic fluxes felt by each sector with $SU(2)_R$ breaking.}
\label{tb:five}
\end{table}
This configuration breaks the $SU(2)_R$ symmetry. 
However, the values of up- and down-type Higgs VEVs 
could not be predicted practically, 
because they are depending on other parts of this model 
and the relevant Higgs phenomenologies, 
which are not the purpose and 
we will not mention it in this paper. 
Since the neutrino Yukawa matrix $y_\nu$ is the transpose 
of up-type quark Yukawa matrix $y_u$ 
and the (Dirac) mass hierarchies among the three generations are the same 
in the two sectors, 
we cannot exchange the magnetic fluxes between up and down sectors. 

In the model shown in Table \ref{tb:model}, the three generations 
are obtained by the magnetic fluxes, 
($M_{ab}=-5$, $M_{bc}=-7$, $M_{ca}=12$) or 
($M_{ab}=-4$, $M_{bc}=-8$, $M_{ca}=12$). 
$Y_{3+4}$, which is defined above Eq.~(\ref{eq:qyuka}), with the 
two types of fluxes have the Gaussian Froggatt-Nielsen texture. 
The matrix $Y_{3+4}$ can be 
approximated by 
\begin{equation}
y_{ij}\sim e^{-\pi({\rm Im}\tau)(a_i+b_j)^2/M_{ab}M_{bc}M_{ca}}
\label{eq:gfnapp}
\end{equation}
with ($a_i=20,~10,~0$) and ($b_i=10,~5,~0$) for 
($M_{ab}=-5$, $M_{bc}=-7$, $M_{ca}=12$), 
and with ($a_i=20,~5,~0$) and ($b_i=15,~5,~0$) 
for ($M_{ab}=-4$, $M_{bc}-8$, $M_{ca}=12$). 
The five-block model shown in Table \ref{tb:five} contains 
a new type of that, that is, ($M_{ab}=-4$, $M_{bc}=-5$, $M_{ca}=9$). 
These fluxes also induce the five generations of Higgs multiplets, 
and then, $Y_{3+4}$ has the Gaussian Froggatt-Nielsen texture and 
can be written by Eq.~(\ref{eq:gfnapp}) 
with ($a_i=10,~5,~0$) and ($b_i=10,~5,~0$).

\end{document}